\documentclass[reprint,twocolumn,secnumarabic,amssymb, nobibnotes, aps, pra,superscriptaddress]{revtex4-1}

\usepackage{extarrows}
\usepackage{amsmath}    
\usepackage{graphicx}    
\usepackage{verbatim}    
\usepackage{color}           
\usepackage{subfigure}   
\usepackage{hyperref}    
\usepackage{verbatim}  
\usepackage{here}
\usepackage{mathrsfs}

\begin{document}
\title{LDA+DMFT approach to resonant inelastic x-ray scattering in correlated materials}

\author{Atsushi Hariki}
\thanks{A.H and M.W contributed equally to this work}
\affiliation{Institute of Solid State Physics, TU Wien, 1040 Vienna, Austria}
\author{Mathias Winder}
\thanks{A.H and M.W contributed equally to this work}
\affiliation{Institute of Solid State Physics, TU Wien, 1040 Vienna, Austria}
\author{Takayuki Uozumi}
\affiliation{Department of Physics and Electronics, Graduate School of Engineering, Osaka Prefecture University 1-1 Gakuen-cho, Nakaku, Sakai, Osaka 599-8531, Japan}
\author{Jan Kune\v{s}}
\affiliation{Institute of Solid State Physics, TU Wien, 1040 Vienna, Austria}
\affiliation{Institute of Physics, Czech Academy of Sciences, Na Slovance 2, 182 21 Praha 8, Czechia}
\date{\today}

\begin{abstract}
    We present a computational study of $L$-edge resonant inelastic x-ray scattering (RIXS) in correlated 3$d$ transition-metal oxides using an {\it ab initio} method based on local density approximation + dynamical mean-field theory (DMFT).
	The present method, building on Anderson impurity model with an optimized continuum bath within DMFT, is an extension of the cluster model to include unbound electron-hole pair excitations as well as material-specific charge-transfer excitations with less empirical parameters.
	We find a good agreement with available experimental data.
	The relationship between correlated bands and fluorescence-like feature in the RIXS spectra is discussed.
\end{abstract}

\maketitle
\section{introduction}
Thanks to a remarkable improvement of its energy resolution in the last decade, resonant inelastic x-ray scattering (RIXS) has become a valued tool for studying materials with strongly correlated electrons~\cite{Ament11}.
It is sensitive to a broad range of excitations from spin, orbital, charge and lattice excitations on the 10-100~meV scale~\cite{Betto17,Fabbris17,Kim12,Braicovich09,Rossi19,Lu18} to atomic-multiplet or charge-transfer (CT) excitations on the eV scale~\cite{Ghiringhelli05,matsubara05,Wang17,Kotani01,groot_kotani}.
Excitations that are not visible to other scattering techniques, such as dipole forbidden excitons, can be observed with RIXS~\cite{Wang18,Kim2014}. 
This comes with a price of complex spectra even including multi-particle excitations, which
makes direct interpretation impractical and
theoretical modeling indispensable.

Numerical simulations of RIXS in solids usually start from either of two limits: the non-interacting solid or the atomic limit. 
The former is based on band theory of effectively non-interacting electrons with the electron-hole excitations
described using Bethe-Salpeter approach~\cite{Vinson11,Gilmore15}.
It provides only a crude approximation of many-body effects in the ground state
as well as in the excited states of correlated materials. 
The latter approach is built around exact diagonalization of the atomic problem and captures the atomic multiplets accurately. Charge transfer to and from the excited transition metal (TM) atom can be incorporated by the cluster model including the nearest-neighbor ligands~\cite{groot_kotani,Haverkort12,Matsubara00} or its extension to multi-site clusters~\cite{Duda06} with more than one TM atom. A rapid growth of the computational cost with the number of sites and orbitals poses a severe limitation on the multi-site extension.

In this paper, we calculate $L$-edge RIXS spectra for a series of TM oxides using an {\it ab initio} approach based on local density approximation (LDA) + dynamical mean-field theory (DMFT)~\cite{metzner89,georges96,kotliar06}.
This approach~\cite{Hariki18,Jindrich18} is a generalization of the cluster model. It allows to include the continuum of unbound electron-hole pairs as well as the CT excitations in a material-specific manner, while retaining the single-impurity description.
To take into account  
the hybridization within valence bands and local electronic correlations, described by LDA+DMFT, as well as the core-valence interaction, the 
Hilbert space of the auxiliary Anderson impurity model (AIM)
is extended by
the core orbitals involved in the RIXS process.
The RIXS spectra are then calculated
with the configuration-interaction impurity solver~\cite{Hariki17}.
This approach not only allows modelling continuum electron-hole excitations, but
eliminates most of the empirical parameters of the traditional cluster model~\cite{Hariki18,Hariki17}.

While the present approach lacks the momentum dependence of bound electron-hole
excitations such as magnons or excitons, it allows a non-perturbative description
of the initial (final) and intermediate states of the RIXS process, the continuum of unbound electron-hole pairs and multi-particle excitations.
Therefore it provides a good description of the incident photon energy $\omega_{\rm in}$-dependence of the RIXS spectra, 
which contains information on electron localization in the intermediate states of RIXS.
A complex situation arises when formation of core-valence excitons compete with continuum excitations in intermediate states at a given $\omega_{\rm in}$~\cite{Hariki18,Pfaff18}.
This is manifested, for example in high-valence nickelates~\cite{Bisogni16} and titanium heterostructures~\cite{Pfaff18}, by  coexistence of Raman-like (RL) and fluorescence-like (FL) features near the x-ray absorption edge.
In this work, we examine the $\omega_{\rm in}$-dependence of the RIXS spectra of 
in NiO, Fe$_2$O$_3$ and cobaltites, typical representatives of correlated 3$d$ TM oxides.

\section{Computational Method}
The computation of $L$-edge RIXS spectra proceeds in two steps.
First, a standard LDA+DMFT calculation is performed as follows.
LDA bands obtained with Wien2K package~\cite{wien2k} are projected~\cite{wien2wannier,wannier90} onto a $dp$ tight-binding model spanning the TM 3$d$ and O 2$p$ orbitals.
The $dp$ model is augmented with the electron-electron interaction within the TM 3$d$ shell.
The on-site Coulomb interaction is parametrized by $U=F^0$ and $J=(F^2+F^4)/14$~\cite{pavarini1,pavarini2}, where $F^0$, $F^2$, and $F^4$ are the Slater integrals~\footnote{We fix the ratio $F^4$/$F^2$=0.625.}. 
$U$ and $J$ values for the studied compounds are given in Sec.~III.
The double-counting term $\mu_{\rm dc}$, which corrects for the $d$--$d$ interaction present in the LDA calculation, renormalizes the bare $p$--$d$ splitting.
While several {\it ad hoc} schemes exist to compute $\mu_{\rm dc}$ 
we treated $\mu_{\rm dc}$ as adjustable parameter fixed by comparison to valence XPS data. 
The strong-coupling continuous-time quantum Monte Carlo impurity solver 
~\cite{werner06,boehnke11,hafermann12,hariki15} was used 
within the self-consistent DMFT calculation. 
After reaching convergence,
the hybridization density $V(\varepsilon)$ is computed on the real
frequency axis 
following the analytic continuation of the self-energy
~\cite{wang09,jarrell96}.

In the second step, we compute $L$-edge RIXS spectra for the AIM with DMFT hybridization function $V(\varepsilon)$ and TM $2p$ core states~\cite{Hariki17}. 
The AIM Hamiltonian $H_{\rm AIM}$ has the form
\begin{equation*}
\hat{H}_{\rm AIM}= \hat{H}_{\rm TM}+ \hat{H}_{\rm hyb}.
\end{equation*}
The on-site Hamiltonian $\hat{H}_{\rm TM}$ is given as
\begin{align}
\hat{H}_{\rm TM}
&=\sum_{\gamma,\sigma}\tilde{\varepsilon}_{d} (\gamma) \hat{d}_{\gamma\sigma}^{\, \dagger} \hat{d}_{\gamma\sigma} 
 + U_{dd} \sum_{\gamma\sigma > \gamma'\sigma'}
  \hat{d}_{\gamma\sigma}^{\, \dagger} \hat{d}_{\gamma\sigma} \hat{d}_{\gamma'\sigma'}^{\, \dagger} \hat{d}_{\gamma'\sigma'}  \notag \\
 &- U_{dc} \sum_{\gamma,\sigma , \,\zeta,\eta} 
  \hat{d}_{\gamma\sigma}^{\, \dagger} \hat{d}_{\gamma\sigma} (1-\hat{c}_{\zeta\eta}^{\, \dagger} \hat{c}_{\zeta\eta}) 
 +\hat{H}_{\rm multiplet}. \notag
\label{eq:Imp_Hamiltonian}
\end{align}
Here, ${\hat{d}_{\gamma\sigma}^{\, \dagger}}$ (${\hat{d}_{\gamma\sigma}}$) and ${\hat{c}_{\zeta\eta}^{\, \dagger}}$ (${\hat{c}_{\zeta\eta}}$) are creation (annihilation) operators  for TM 3$d$ and 2$p$ electrons, respectively.
The ${\gamma}$ (${\zeta}$) and ${\sigma}$ (${\eta}$) are TM ${3d}$ (2$p$) orbital and spin indices.
The TM $3d$ site energies are given as $\tilde\varepsilon_{d}(\gamma)=\varepsilon_{d}(\gamma)-\mu_{dc}$, where $\varepsilon_d(\gamma)$ are the energies of the Wannier states and $\mu_{dc}$ is the double-counting term mentioned above. 
The isotropic part of the 3$d$ -- 3$d$ ($U_{dd}$) and 2$p$ -- 3$d$ ($U_{dc}$) interactions are shown explicitly,
while terms containing higher Slater integrals and the spin-orbit interaction are contained in the $\hat{H}_{\rm multiplet}$ term.
The spin-orbit coupling within the TM 2$p$ and 3$d$ shell and the anisotropic part of the 2$p$-3$d$ interaction parameters $F^k$, $G^k$ are calculated with an atomic Hartree-Fock code~\cite{cowan}. 
The computed values of $F^k$ and $G^k$ are scaled by 80\%
~\footnote{This simulates the effect of intra-atomic configuration interaction 
from higher basis configurations neglected in the atomic calculation} and we fix the isotropic part of the core-valence interaction by the
empirical relation $U_{dc}=1.2 \times U_{dd}$
~\cite{groot_kotani,cowan,Sugar72,matsubara05}.
The $\hat{H}_{\rm hyb}$ term describes hybridization with the fermionic bath 
\begin{equation}
\hat{H}_{\rm hyb} = \sum_{\alpha,\gamma\sigma} \epsilon_{\alpha,\gamma\sigma} \hat{v}_{\alpha,\gamma\sigma}^\dagger \hat{v}_{\alpha,\gamma\sigma} + \sum_{\alpha,\gamma\sigma} V_{\alpha,\gamma\sigma}(\hat{d}_{\gamma\sigma}^\dagger \hat{v}_{\alpha,\gamma\sigma} + h.c ).\notag
\end{equation}
The first term represents the energies of the auxiliary orbitals and the second term 
describes the hopping between the TM 3$d$ state and the auxiliary orbitals with the amplitude $V_{\alpha,\gamma\sigma}$.
Here, ${\hat{v}_{\alpha,\gamma\sigma}^{\, \dagger}}$ (${\hat{v}_{\alpha,\gamma\sigma}}$) is the creation (annihilation) operator for the auxiliary state with energy $\epsilon_{\alpha,\gamma\sigma}$.
The amplitude $V_{\alpha,\gamma\sigma}$ relates to the DMFT hybridization density $V_{\gamma\sigma}^2(\varepsilon)$ by
\begin{equation*}
V_{\gamma\sigma}^2(\varepsilon)=-\frac{1}{\pi}\operatorname{Im}\sum_{\alpha}\frac{V^2_{\alpha,\gamma\sigma}}{\varepsilon-\varepsilon_{\alpha,\gamma\sigma}}.
\end{equation*}
The $V_{\gamma\sigma}^2(\varepsilon)$ encodes the information about electron hopping between a given TM orbital $\gamma$ (spin $\sigma$) and the rest of the crystal~\cite{kotliar06,Hariki17}.
In practice, $V_{\gamma\sigma}^2(\varepsilon)$ obtained with the LDA+DMFT calculation is represented by 25 discretized bath states $\alpha$ for each orbital $\gamma$ and $\sigma$~\footnote{The hybridization is assumed to be orbital (and spin) diagonal, which is a good approximation in the studied compounds.}.
The RIXS intensity at finite temperature $T$ is given
by~\cite{groot_kotani,matsubara05,Kramers25}
\begin{align}
\label{eq:rixs}
F^{(n)}_{\rm RIXS}(\omega_{\rm out},\omega_{\rm in})&=\sum_{f} \left| \sum_{m}
     \frac{\langle f | T_{\rm e} | m\rangle \langle m | T_{\rm i} | n \rangle }
     {\omega_{\rm in}+E_n-E_m+i\Gamma}
      \right|^2 \notag \\ 
   &\times \delta(\omega_{\rm in}+E_n-\omega_{\rm out}-E_f) \\
   =&\sum_{f} \left| 
     \langle f | T_{\rm e}
     \frac{1}{\omega_{\rm in}+E_n-H_{\rm AIM}+i\Gamma}  T_{\rm i} | n \rangle
      \right|^2 \notag \\
     &\times  \delta(\omega_{\rm in}+E_n-\omega_{\rm out}-E_f). 
\end{align}
Here, $|n\rangle$, $|m\rangle$, and $|f\rangle$ represent the initial, intermediate, and final states with energies $E_n$, $E_m$, and $E_f$, respectively.
The individual contributions from the initial states are averaged over, weighted with the Boltzmann factors~\cite{Hariki17,Hariki18}.
$\Gamma$ is the inverse lifetime of the core-hole in the intermediate state, set to 300~meV throughout the present study.
$T_i$ ($T_e$) is the transition operator that describes the x-ray absorption (emission) in the RIXS process
and encodes the experimental geometry~\cite{Matsubara00}.
In the present study, we use a setting, in which the polarization of the x-rays is perpendicular (parallel) to the scattering plane for NiO, Fe$_2$O$_3$ (cobaltities)~\footnote{We take average of RIXS intensities calculated for two independent polarizations of the emitted x-rays since the polarization of the emitted x-rays is not resolved in standard RIXS measurements. }. The scattering angle is set to $90^\circ$ with the grazing angle of $20^\circ$ for the incident x-rays, simulating a typical experimental setup.  
The incident (emitted) x-ray has the energy $\omega_{\rm in}$  ($\omega_{\rm out}$)
and energy loss is given by $\omega_{\rm loss}=\omega_{\rm in}-\omega_{\rm out}$.
The configuration interaction scheme is employed to compute the RIXS intensity for the AIM~\cite{Hariki17,Hariki18}.
The initial states are computed using the Lanczos method. 
Their propagation by the resolvent $(\omega_{\rm in}+E_n-H_{\rm AIM}+i\Gamma)^{-1}T_{i}|n\rangle$ is computed using conjugate-gradient-based method. 
Though the RIXS calculation for different photon energies $\omega_{\rm in}$ can be parallelized in a straightforward way, one can also adopt the shift and seed-switching techniques in the conjugate-gradient-based method (for constant $\Gamma$ case)~\cite{Yamamoto08,Sogabe07}, see Appendix.~I.

\begin{figure*}[t] 
\includegraphics[width=175mm]{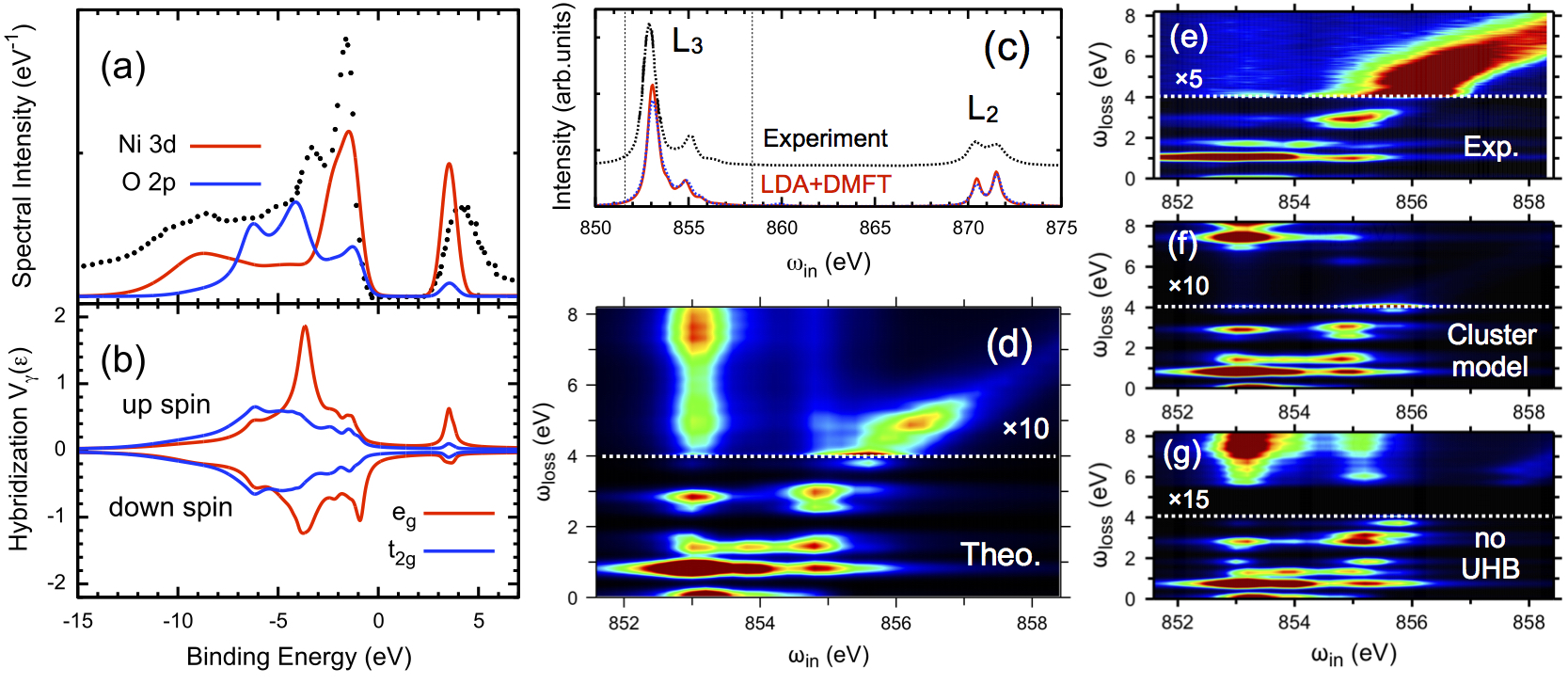}
\caption{(a) LDA+DMFT valence spectra of NiO. The experimental data (black, dotted) are taken from Ref.~\onlinecite{Sawatzky84}. (b) DMFT hybridization function. (c) Ni $L_{3}$-edge XAS calculated by LDA+DMFT (solid), cluster model (blue, dashed) and the experimental data in Ref.~\onlinecite{Alders98}.
RIXS spectra calculated by (d) LDA+DMFT. 
(e) experimental data~\cite{Ghiringhelli09_nio}.
(f) RIXS spectra calculated by the cluster model. 
(g) RIXS spectra calculated without hybridization intensities from $-2.0$ to $0.0$~eV. 
The RIXS intensities above the horizontal lines (white) are magnified by a factor indicated in panels.
The spectral broadening is taken into account using a Gaussian of 150~meV for RIXS, a Lorentzian 300~meV for XAS, and a Gaussian 600~meV for valence XPS.}
\label{fig_nio}
\end{figure*}

The x-ray absorption spectroscopy (XAS) is simulated with the same AIM as RIXS.
The XAS final states are the intermediate states of the RIXS process.
The XAS spectra are therefore closely related to the 
$\omega_{\rm in}$-dependence of the RIXS intensities.
The contribution to XAS from an initial state $|n\rangle$ is given by
\begin{equation}
\label{eq:xas}
\begin{split}
F^{(n)}_{\rm XAS}(\omega_{\rm in})&=-\frac{1}{\pi}\operatorname{Im}  \sum_{n} \langle n | T^{\dag}_{\rm i} \frac{1}{\omega_{\rm in}+E_n-H_{\rm AIM}+i\Gamma} T_{\rm i}| n \rangle \notag. \\
\end{split}
\end{equation}

For comparison, we present $L$-edge XAS and RIXS spectra calculated by the cluster model.
The on-site Hamiltonian of the cluster model has the same form as $\hat{H}_{\rm TM}$, while the hybridization part takes into account only molecular orbitals composed of nearest-neighboring ligand $p$ states, thus inevitably excitations are bounded within the cluster. Our construction of the cluster model can be found in Ref.~\onlinecite{Ghiasi19}.

\section{results and discussion}
\subsection{NiO}

Fig.~\ref{fig_nio}a shows the valence spectra of NiO calculated by LDA+DMFT
in the antiferromagnetic state at $T=300$~K (below the experimental N\'eel temperature of 525~K).
We employed $U=7.0$~eV and $J=1.1$~eV~\cite{Hariki17}.
We find a fair agreement with experimental photoemission and inverse photoemission data~\cite{Sawatzky84} 
for $\mu_{\rm dc}$ in the range of $50-52$~eV
(The $\mu_{\rm dc}$-dependence of valence, XAS and RIXS spectra can be found in Appendix.~II.).
Here we present the result obtained with $\mu_{\rm dc}=50$~eV.
Fig.~\ref{fig_nio}c shows Ni $L_{2,3}$-edge XAS calculated using the LDA+DMFT and cluster model, together with the experimental data~\cite{Alders98}.
The Ni $L_{2,3}$ XAS is composed of the main line ($\omega_{\rm in}$ between $850-855$~eV), corresponding to $|cd^9\rangle$ final-state configuration, and the weak satellite ($\omega_{\rm in} \sim 856$~eV), corresponding to $|\underline{c}d^{10}\underline{v}\rangle$ configuration.
Here, $\underline{c}$ and $\underline{v}$ denote a hole in 2$p$ core level and valence bands, respectively.
The LDA+DMFT and cluster-model results are almost identical to each other and show a good agreement with the experimental data. The match of the two is expected as
the CT screening from the surrounding atoms is rather weak in the XAS final states.

Fig.~\ref{fig_nio}d shows Ni $L_3$-RIXS map obtained by LDA+DMFT.
For comparison, Figs.~\ref{fig_nio}ef show the cluster-model result and the experimental data~\cite{Ghiringhelli09_nio}.
Three distinct RIXS features are observed: RL $d$--$d$ excitations ($\omega_{\rm loss}=1-4$~eV); CT excitations ($\omega_{\rm loss}=4-8$~eV) showing a broad-band feature along $\omega_{\rm loss}$; FL feature, showing a linear increasing feature with $\omega_{\rm in}$.
The RL and CT excitations resonate mainly at the $L_3$ main line, while the FL feature appears for $\omega_{\rm in}>$~855~eV. 
The LDA+DMFT result 
shows a good overall agreement with the experimental data. 
In the cluster-model result, though the RL feature is reproduced, the CT feature is found at a sharp $\omega_{\rm loss}$ and the FL feature is missing due to the lack of the unbound electron-hole pair (EHP) continuum in this description.
The lowest $d$--$d$ peak at 1.0~eV in the experimental data, corresponding to a single excitation from $t_{2g}$ orbit to $e_{g}$ orbit in the one-electron picture~\footnote{In terms of the atomic symbol, this corresponds to $ {}^3{A}_{2} \rightarrow {}^3{T}_2$ excitation.}, is located at around 0.85~eV in both the LDA+DMFT and cluster-model results, see also Appendix.~II.
The quantitative discrepancy could be attributed to underestimation of the $e_{g}$--$t_{2g}$ splitting due to covalency in the present LDA calculation~\cite{Haverkort12}.


\begin{figure*}[ht] 
\includegraphics[width=175mm]{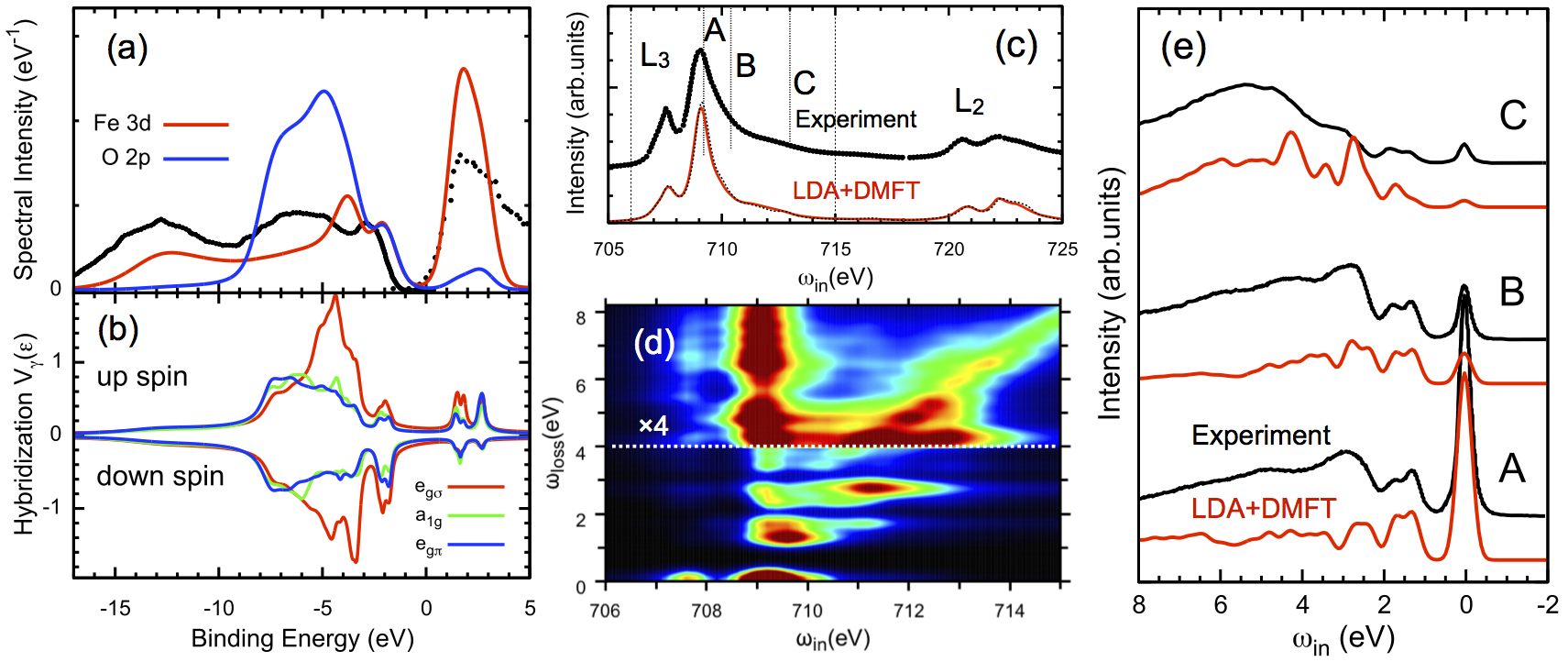}
\caption{(a) LDA+DMFT valence spectra of Fe$_2$O$_3$ with the experimental data (black)~\cite{Ciccacci91,Lad89}.
(b) DMFT hybridization function $V(\varepsilon)$.
(c) Fe $L_{2,3}$-edge XAS spectra calculated by LDA+DMFT (solid), cluster model (dashed) and experimental data (dotted)~\cite{Yang09}.
(d) RIXS spectra calculated by LDA+DMFT. 
The intensities above the horizontal lines (white) are magnified by the factor indicated in panels.
(e) RIXS spectra calculated at selected incident photon energies, see panel (c).
The experimental data are taken from Ref.~\onlinecite{Miyawaki17} (A-C corresponds to 3, 5, 7 in the reference). 
The spectral broadening is taken into account using a Gaussian of 200~meV for RIXS, a Lorentzian 300~meV for XAS, and a Gaussian 600~meV for valence XPS.}
\label{fig_fe2o3}
\end{figure*}

The FL feature originates from unbound EHP excitations.
The low $\omega_{\rm loss}$-region of the FL features reflects the EHPs that involve low-energy valence bands, as demonstrated in Fig.~\ref{fig_nio}g. 
There the hybridization intensities $V(\varepsilon)$ (from $-2$ to $0$~eV), see Fig.~\ref{fig_nio}b, is numerically removed, that forbids residence of a hole in the low-energy valence bands in the RIXS process.
This eliminates the low-$\omega_{\rm loss}$ FL feature around $4-6$~eV.

Finally we comment on a character of the FL RIXS feature in a large-gap insulator.
In Appendix.~III, we show the RIXS spectra calculated while excluding a CT from x-ray excited Ni ion to the conduction bands above Fermi energy $E_F$, that forbids the excitation of UH states ($d^9$) outside the excited Ni ion in the RIXS process.
This results in only a minor intensity modulation of the FL feature, suggesting that the FL $L_3$-RIXS feature of NiO reflects projected EHP continuum with an extra $d$ electron sitting on the excited Ni site (local UH state) and a hole propagating in the LH or O 2$p$ bands.
This observation would be common  in a large-gap system and qualitatively differs from the behavior of the FL feature in high-valence TMO with a small gap~\cite{Hariki18}.


\subsection{Fe$_2$O$_3$}

Fig.~\ref{fig_fe2o3}a shows the valence spectra of Fe$_2$O$_3$ obtained by LDA+DMFT in the experimental corundum structure~\cite{Finger80} and antiferromagnetic state at $T=300$~K (the experimental N\'eel temperature is 950~K).
We employ $U=6.8$~eV and $J=0.86$~eV following previous DFT studies~\cite{Anisimov91,Kunes09b}.
A reasonable agreement with experimental photoemission and inverse photoemission data~\cite{Ciccacci91,Lad89} is found in the range $\mu_{\rm dc}=30.6-32.6$~eV.
Thus we present the result obtained with $\mu_{\rm dc}=31.6$~eV.
The $\mu_{\rm dc}$-dependence of valence, XAS and RIXS spectra can be found in Appendix.~II.
The hybridization density in Fig.~\ref{fig_fe2o3}b shows the spin dependence reflecting the antiferromagnetic ordering.
Fig.~\ref{fig_fe2o3}c shows Fe $L_{2,3}$-edge XAS calculated by LDA+DMFT and the cluster model, together with the experimental data~\cite{Yang09}.
The two methods yield almost identical results and show a good agreement with the experiment.
The shape of the Fe $L_{3}$-edge main line ($706$ -- $711$~eV), that corresponds to the $|\underline{c}d^6\rangle$ final state, is known to be sensitive to the local multiplet structure~\cite{Groot05,Groot90}, indicating the accuracy of the parameters in the present local Hamiltonian $\hat{H}_{\rm TM}$.   

Fig.~\ref{fig_fe2o3}d shows Fe $L_3$-RIXS map obtained by the LDA+DMFT approach.
The RIXS intensities calculated at selected photon energies are shown in Fig.~\ref{fig_fe2o3}e with recent high-resolution experimental data~\cite{Miyawaki17}.
Fe $L_3$ RIXS shows rich $d$--$d$ features ($\omega_{\rm loss}=1$ -- $5$~eV) and a complex $\omega_{\rm in}$ dependence due to a variety of multiplets in the $d^5$ manifolds.
The LDA+DMFT result reproduces the position and $\omega_{\rm in}$-dependence of low-energy features reasonably well.

\begin{figure*}[t] 
\includegraphics[width=180mm]{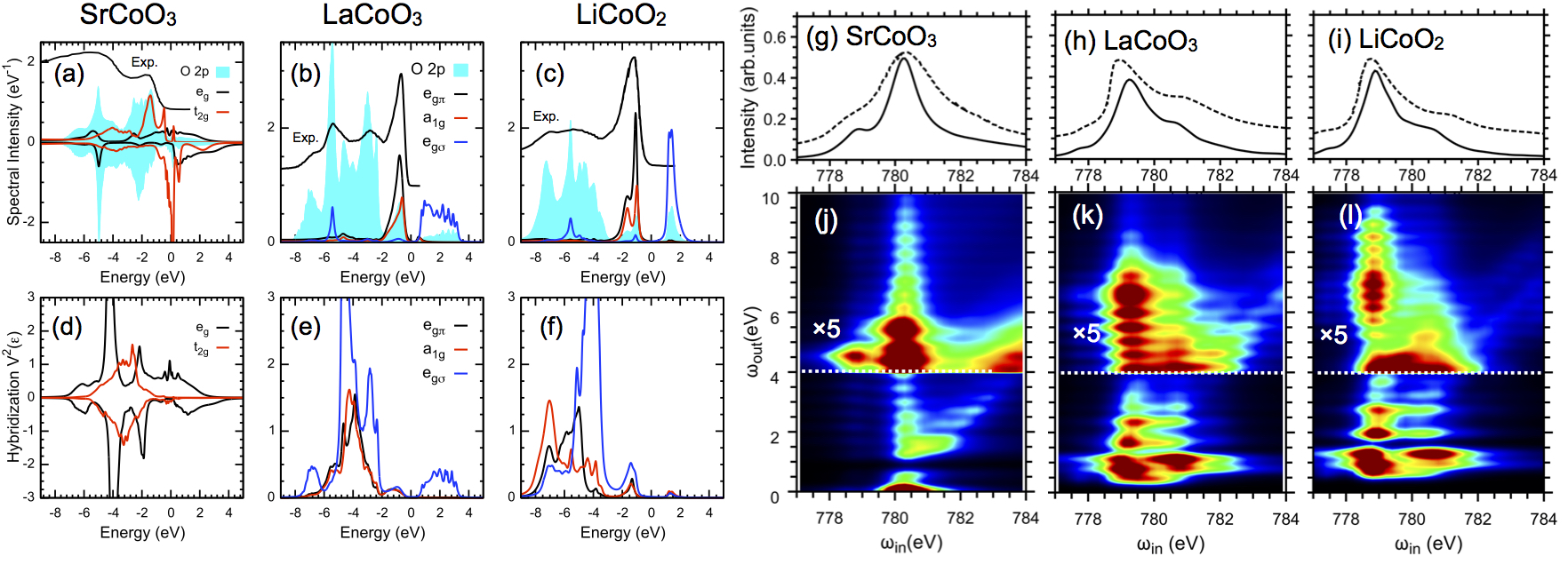}
\caption{LDA+DMFT Valence spectra and hybridization intensities of (a,d) SrCoO$_3$, (b,e) LaCoO$_3$ and (c,f) LiCoO$_2$.
Co $t_{2g}$ orbitals split into $e_{g \pi}$ and $a_{1g}$ orbitals due to trigonal distortion in LaCoO$_3$ and LiCoO$_2$.
The experimental valence photoemission data for SrCoO$_{3-\delta}$~\cite{Zhao19}, LaCoO$_3$ (Sr 0.2\% doped)~\cite{Takegami19} and LiCoO$_2$~\cite{Galakhov06} are shown together.
The Co $L_3$-edge XAS and RIXS spectra calculated for (g,j) SrCoO$_3$, (h,k) LaCoO$_2$ and (i,l) LiCoO$_2$, together with the experimental XAS data (dashed lines)~\cite{Haverkort06,Jiali19,Mizokawa13}. 
The RIXS intensities above horizontal lines (white) are magnified by a factor indicated in panels.
The spectral broadening is taken into account using a Gaussian of 150~meV for RIXS, a Lorentzian 300~meV for XAS.}
\label{fig_co}
\end{figure*}


\subsection{ Cobaltites }

We present Co $L$-edge RIXS spectra in representative cobaltites; SrCoO$_3$, LaCoO$_3$ and LiCoO$_2$.
The comparison among the three enables to explore how RIXS features vary with valency or lattice geometry.
SrCoO$_3$ and LaCoO$_3$ crystallize in the corner-sharing perovskite structure, while LiCoO$_2$ crystallizes in a quasi-two-dimensional structure with edge-sharing CoO$_6$ octahedra.
Formally Co ion is trivalent ($3d^6$) in LaCoO$_3$ and LiCoO$_2$, while it is tetravalent ($3d^5$) in SrCoO$_3$.
Due to its small CT energy, SrCoO$_3$ possesses a dominant $d^6$ configuration (plus one hole in ligands) in the ground state~\cite{Kunes12,Saitoh97,Potze95}.
The Co $d^6$ manifolds have rich low-energy multiplets characterized by low-spin ($S=0$, LS), intermediate-spin ($S=1$, IS) and high-spin ($S=2$, HS) states.
The ground states of the three compounds at low-temperatures are well known; LaCoO$_3$ and LiCoO$_2$ are band insulators (insulating gap $\sim 0.5$~eV) with the LS configuration, while SrCoO$_3$ is a ferromagnetic metal with an admixture of the HS state and charge fluctuations around it~\cite{Kunes12}.
Note that some of the present authors reported the (bound) excitonic dispersion of the IS state in $L_3$-edge RIXS spectrum of LaCoO$_3$~\cite{Wang18}, which cannot be captured in the present AIM approach and thus is out of the scope of this study.
The LDA+DMFT calculations are performed in the experimental crystal structure reported well below possible spin-state transition temperatures.
Following previous DFT studies for LaCoO$_3$~\cite{Krapek12}, we use $U=6.0$~eV and $J=0.8$~eV.

Figs.~\ref{fig_co}abc show the LDA+DMFT valence spectra, together with experimental data.
Due to its LS character, $t_{2g}$ states are almost fully occupied in LaCoO$_3$ and LiCoO$_2$, while the HS character in SrCoO$_3$ yields considerable $e_g$ weights below $E_F$ in the majority-spin channel~\cite{Kunes12}.
Figs.~\ref{fig_co}def show the hybridization intensities $V^2(\varepsilon)$.
The intensities around $-8$ to $-2$~eV ($-2$ to $4$~eV) represent the hybridization with O 2$p$ (Co 3$d$) states though explicit decomposition of contributing states in the continuum bath is impossible~\cite{Ghiasi19}. 
Despite the similar LS valence spectra in LiCoO$_2$ and LaCoO$_3$, we find a clear difference in $V^2(\varepsilon)$ for the $e_{g}$ orbital between the two.
LaCoO$_3$ shows sizable hybridization intensities above $E_F$, while LiCoO$_2$ shows only below $E_F$ (around $-2$~eV).
In LaCoO$_3$ with nearly 180$^\circ$ of Co-O-Co bonds, inter-orbital ($e_g$-$t_{2g}$ channel) hopping between neighboring Co sites is forbidden, while it is allowed in LiCoO$_2$ owing to the edge-sharing CoO$_6$ octahedra.
The $e_g$-$e_{g}$ hopping, on the other hand, is allowed/forbidden in the former/latter geometry.
This explains the presence/absence of the hybridization intensities with the empty $e_g$ bands above $E_F$ in LaCoO$_3$/LiCoO$_2$.
In this way, $V^2(\varepsilon)$ encodes the lattice environment around the impurity site. 
Since an extra $d$ electron, excited by the local x-ray absorption, goes into the empty $e_g$ states in the LS configuration, the hybridization properties of $e_g$ orbital is important to understand possible EHP excitations in the RIXS spectra.

Figs.~\ref{fig_co}ghi show the Co $L_3$-XAS spectra calculated by LDA+DMFT.
In both trivalent~\cite{Haverkort06} and tetravalent cases~\cite{Potze95}, the Co $L_3$-XAS is sensitive to the spin-state character on the Co atom in the ground state.
Thus the overall good agreement with the available experimental data~\cite{Haverkort06,Jiali19,Mizokawa13} suggests that the spin state in the ground state is well described within the LDA+DMFT scheme.

Figs.~\ref{fig_co}jkl show the RIXS spectra calculated across the Co $L_3$ edge.
The $d$--$d$ excitations in LaCoO$_3$ and LiCoO$_2$ resemble each other due to the similar local multiplet structures above the LS ground state, while those in SrCoO$_3$ are rather obscure mainly due to the thermal mixture of the HS multiplets.
Despite the similarity of the $d$--$d$ excitations, the FL feature in LaCoO$_3$/LiCoO$_2$ is visible/invisible.
This difference originates from the hybridization of the excited Co ion with the continuum of conduction states above $E_F$, that differs in the two lattice geometries as mentioned above. 
The presence/absence of the FL feature in the corner/edge sharing structure resembles the behavior of the FL feature isoelectronic high-valence cuprates (LaCuO$_3$ and NaCuO$_2$), theoretically predicted recently~\cite{Hariki18}.
The FL feature in SrCoO$_3$ is more intense compared to that in LaCoO$_3$ despite comparable hybridization intensities above $E_F$ between the two, see Figs.~\ref{fig_co}de.
This is because, in SrCoO$_3$, metallicity due to negative CT energy favors EHP excitations.

\section{Conclusion}

We presented numerical simulations of $L$-edge RIXS spectra of typical 3$d$ transition-metal oxides:~NiO, Fe$_2$O$_3$ and cobaltites obtained using LDA+DMFT approach.
The present method is based on the Anderson impurity model with the DMFT continuum bath augmented by the relevant core states. It provides an extension of the cluster model to include unbound EHP excitation as well as the CT excitation in material-specific manner. 
The approach reproduces well the experimental RIXS and XAS data of
the studied materials which includes NiO, Fe$_2$O$_3$ and several cobaltites.
Taking cobaltities as an example, we examined the change of RIXS features with valency or crystal geometry. We found substantial differences in RIXS spectra of isoelectronic 
LaCoO$_3$ and LiCoO$_2$ despite their almost identical valence photoemission and XAS spectra. The difference between the two compounds lies in the decoration of the crystal lattice with CoO$_6$ octahedra, which is encoded the DMFT hybridization function.
This example demonstrates that the information contained in the RIXS spectra
cannot be extracted from one-particle spectral function, e.g. by convolution.

The present method provides computationally-feasible material specific approach
to RIXS spectra in wide range of materials including the strongly correlated ones.

\begin{acknowledgments}
The authors thank G. Ghiringhelli for providing his experimental data of NiO and for valuable discussions. 
We thank A. Sotnikov and J. Fern\'andez Afonso for fruitful discussions.
A.H., M.W., and J.K. are supported by the European Research Council (ERC)
under the European Union's Horizon 2020 research and innovation programme (Grant Agreement No.~646807-EXMAG).
T.U. was supported by JSPS KAKENHI Grant Number JP16K05407. 
The computational calculations were performed at the Vienna Scientific Cluster (VSC).
\end{acknowledgments}

\begin{figure}[t] 
\includegraphics[width=85mm]{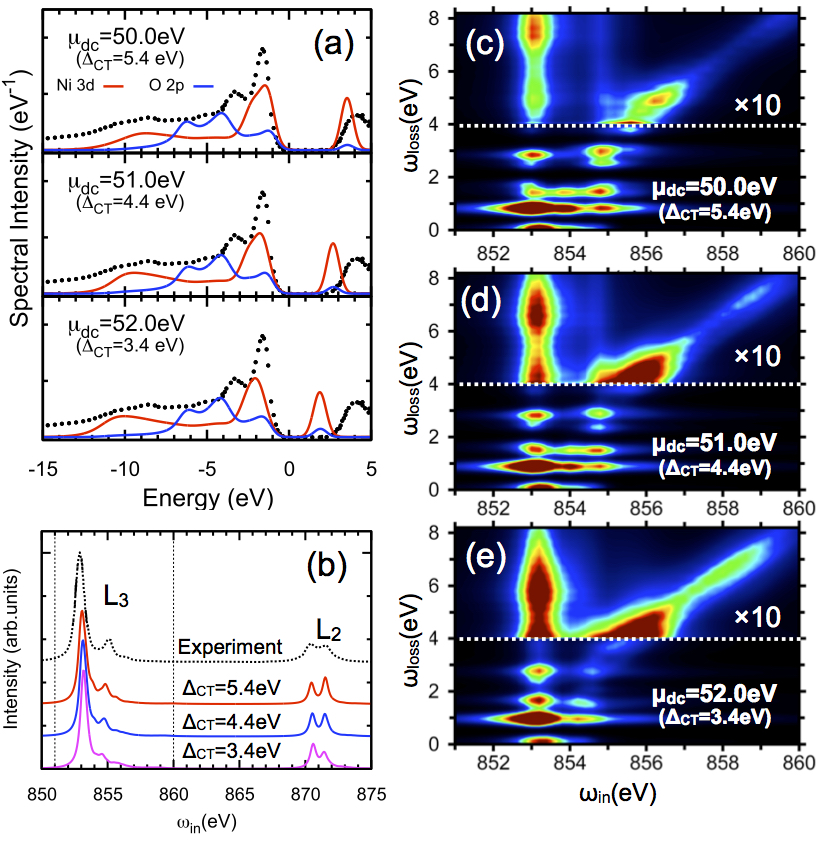}
\caption{The double-counting correction $\mu_{\rm dc}$ dependence of (a) valence spectra, (b) Ni $L_{2,3}$-XAS spectra and (c-e) Ni $L_3$-RIXS spectra of NiO calculated by LDA+DMFT.
The RIXS intensities above the horizontal lines (white) are magnified by a factor indicated in panels.}
\label{fig_nioapp}
\end{figure}


\section*{\label{app3} Appendix.~I:~Shifted Conjugate-Gradient method in RIXS calculation }

Here we introduce numerical methods for computing RIXS intensities.
The initial states $|n\rangle$, that contribute to thermal average at the simulated temperatures, are calculated using the Lanczos method.
The complete spectrum of the intermediate states $\{|m\rangle \}$ in Eq.~(1) is usually not available for a large Hamiltonian.
As seen in Eq.~(2), however for computing RIXS intensities, one only needs propagated vectors $|x_n (\omega_{\rm in})\rangle=(\omega_{\rm in}+E_n-H_{\rm AIM}+i\Gamma)^{-1}T_{i}|n\rangle$.
To obtain the $|x_n (\omega_{\rm in})\rangle$ vectors, the (high dimensional) linear equations are solved using the conjugate-gradient-based (CG) method.
Note that, because of the presence of the (inverse) lifetime term $i\Gamma$, one should adopt conjugate-orthogonal CG (COCG)~\cite{Vorst90} method for real $H_{\rm AIM}$ (i.e.~$H_{\rm AIM}-i\Gamma$ is not Hermite but symmetric) and use bi-conjugate gradient (BiCG) method for complex $H_{\rm AIM}$ (i.e.~$H_{\rm AIM}-i\Gamma$ is neither Hermite nor symmetric). 
Though the CG method searches for the solution of the linear equation above with (sparse) large $H_{\rm AIM}$ in an efficient way, the 
most computationally demanding part is the iterative matrix-vector product in the subspace construction.
A straightforward way is to parallelize the intermediate-state calculation for different photon energies $\omega_{\rm in}$.
Another route is to use the so-called shifted CG technique~\cite{Yamamoto08,Sogabe07} that builds on the (scalar) shift invariance of the Krylov subspace with fixed starting vector ($|n\rangle$)
\begin{equation*}
{\mathscr K}_k (\omega I+h_n, |n\rangle)={\mathscr K}_k (h_n, |n\rangle),
\end{equation*}
where $h_n=\omega_{\rm ref}+E_n-H_{\rm AIM}+i\Gamma$ and $\omega_{\rm ref}$ is a reference photon energy.
The ${\mathscr K}_l$ denotes the Krylov subspace with $k$-th order defined as
\begin{equation*}
{\mathscr K}_k (h_n, |n\rangle):={\rm span} \{ |n\rangle, h_n|n\rangle, h_n^2|n\rangle, \cdots, h_n^{k-1}|n\rangle  \}.
\end{equation*}
Using the shift invariance property of the Krylov subspace, one can solve the COCG/BiCG recursion formula for the target photon energy $\omega_{\rm in}$ (appear via $\omega=\omega_{\rm in}-\omega_{\rm ref}$) without any matrix-vector products, see Refs.~\onlinecite{Yamamoto08,Sogabe07} for shifted COCG and Ref.~\onlinecite{Frommer2003} for shifted BiCG and its variants.
The main advantage of the shift technique over a brute parallelization over photon energies is saving the computational sources/memory, perhaps being an issue for huge $H_{\rm AIM}$ or dense $\omega_{\rm in}$ mesh.
However a tricky issue in the shift technique is that one may need a prior knowledge for the dimensions of the Krylov subspace necessary for achieving the converged solution for all photon energies.
In $L_3$-edge RIXS calculations, the convergence usually depends strongly on the photon energies $\omega_{\rm in}$;~the convergence for localized intermediate states (e.g.~near the $L_3$ main edge) is rather fast, while that for the continuum ones (e.g.~far above the main edge) sometime requires 100--1000 iterations.
In practice, we recommend that one starts the calculation with the highest photon energy (far above the target edge) as a reference energy $\omega_{\rm ref}$ and subsequently approaches to the main edge using the shifted technique. 
When further expansion of the Krylov subspace is necessary, one could use the seed switching technique~\cite{Yamamoto08,Sogabe07}, that avoids restarting the subspace construction for a new photon energy. 

\begin{figure}[t] 
\includegraphics[width=85mm]{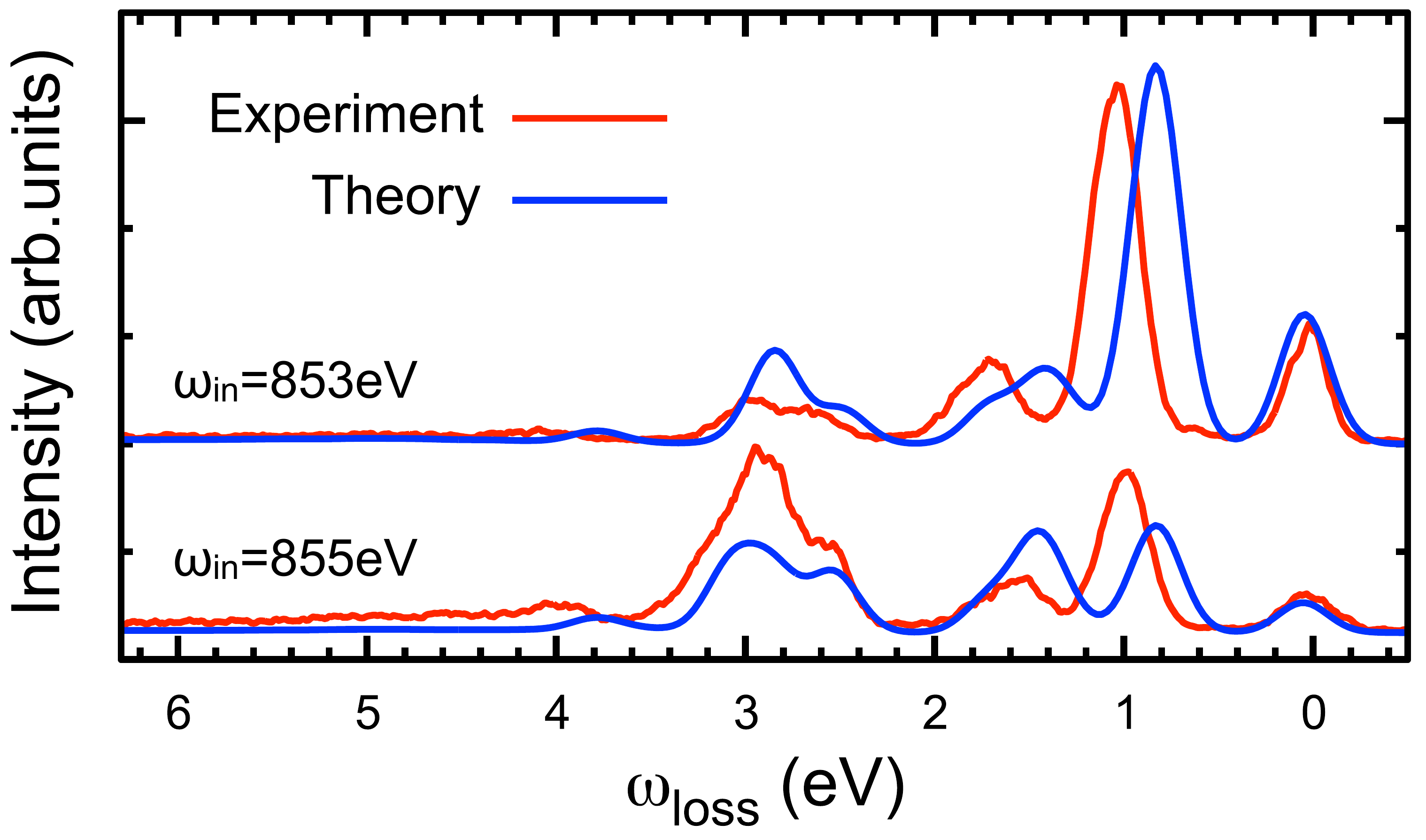}
\caption{Ni $L_3$-RIXS intensities calculated by LDA+DMFT ($\mu_{\rm dc}$=50~eV) for selected $\omega_{\rm in}$. The experimental data are taken from Ref.~\onlinecite{Ghiringhelli09_nio}.}
\label{fig_nioapp2}
\end{figure}

\section*{\label{app2} Appendix.~II:~Double-counting correction dependence }

\begin{figure}[th] 
\includegraphics[width=85mm]{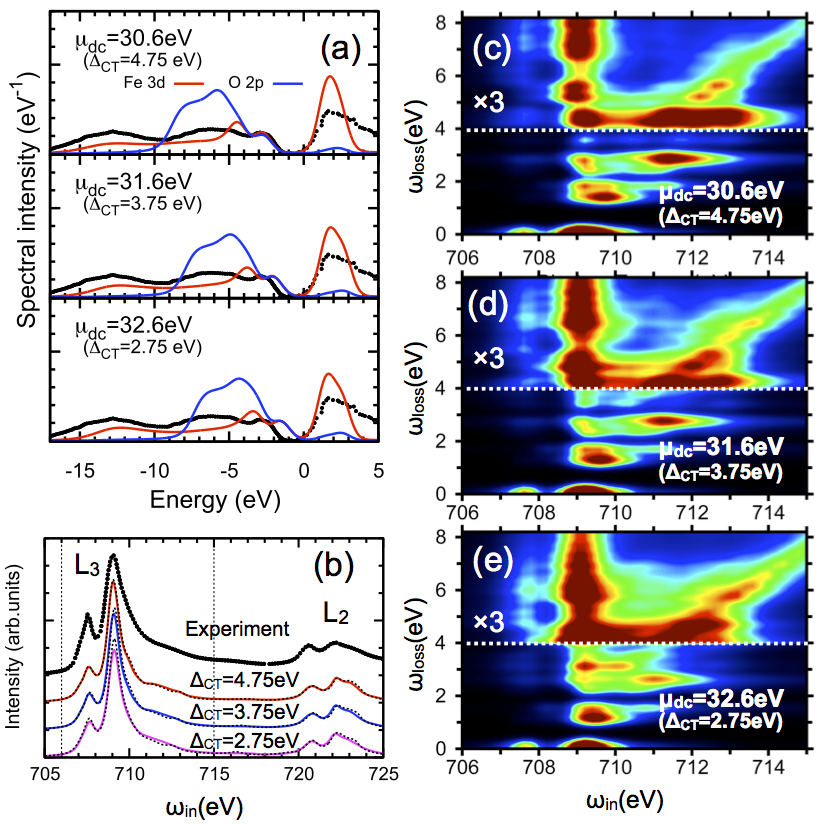}
\caption{The double-counting correction $\mu_{\rm dc}$ dependence of (a) valence spectra, (b) Fe $L_{2,3}$-XAS spectra and (c-e) Fe $L_3$-RIXS spectra of Fe$_2$O$_3$ calculated by LDA+DMFT.
The RIXS intensities above the horizontal lines (white) are magnified by a factor indicated in panels.}
\label{fig_fe2o3app}
\end{figure}

Figures~\ref{fig_nioapp} and \ref{fig_fe2o3app} summarize the $\mu_{\rm dc}$-dependence of the LDA+DMFT result for valence, XAS and RIXS spectra in NiO and Fe$_2$O$_3$, respectively.

In NiO, the one-particle gap reduces with $\mu_{\rm dc}$ increase (corresponding to decrease of the CT energy $\Delta_{\rm CT}$), as expected in the CT-type insulator~\cite{Zaanen85}.
The satellite and lower Hubbard band are observed around 9~eV and 1~eV, respectively~\footnote{In terms of configurations, the satellite corresponds to $d^7$ final states and $d^8v$ in the photoemission final states.}.
We obtained a reasonable agreement with the experimental photoemission and inverse photoemission data~\cite{Sawatzky84} in the range of $\mu_{\rm dc}=50-52$~eV.
The $\mu_{\rm dc}$-dependence of the Ni $L_{2,3}$-XAS spectra is rather weak since the spectral shape is mostly dominated by the local multiplet interaction and the crystal-field splitting. 
The onset of the FL feature in the $\omega_{\rm in}$-$\omega_{\rm loss}$ plot relates to the one-particle gap in the valence spectra.
Fig.~\ref{fig_nioapp2} shows the RIXS intensities calculated by LDA+DMFT for selected photon energies $\omega_{\rm in}$, together with the experimental data~\cite{Ghiringhelli09_nio}.

In Fe$_2$O$_3$, a reasonable agreement with experimental photoemission and inverse photoemission data~\cite{Ciccacci91,Lad89} is found in the range of $\mu_{\rm dc}=30.6-32.6$~eV.
The Fe $L_{2,3}$-edge XAS spectra are rather insensitive to the choice of the double-counting corrections, indicating the spectral features are dominated by the local multiplets~\cite{Groot05,Groot90}.

\section*{\label{app1} Appendix.~III:~with/without upper Hubbard band }
\begin{figure}[h] 
\includegraphics[width=85mm]{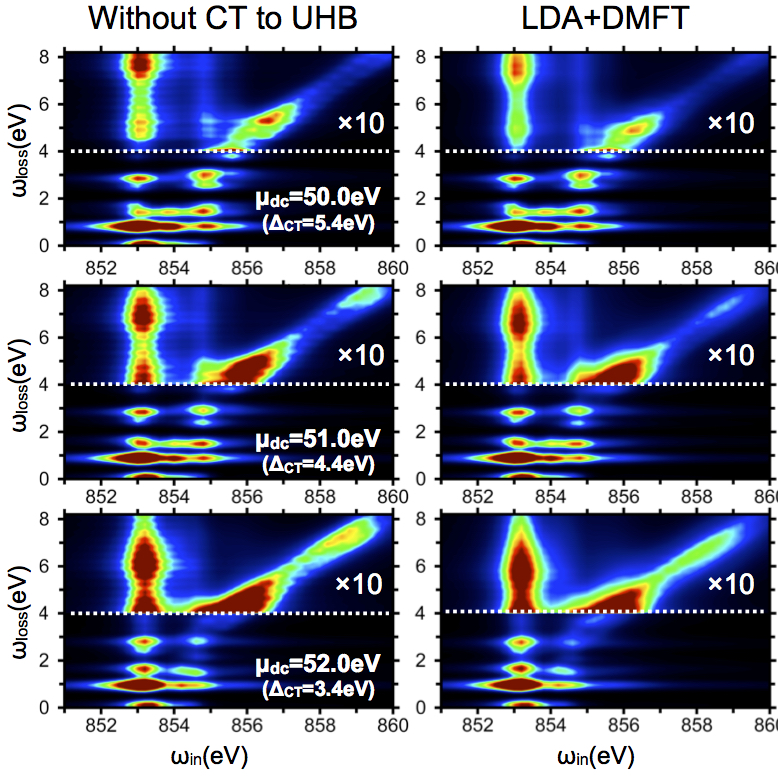}
\caption{ (Left) Ni $L_3$-RIXS map of NiO calculated by excluding a CT between x-ray excited Ni ion and conduction states above Fermi energy $E_F$. (Right) Ni $L_3$-RIXS map of NiO calculated by LDA+DMFT.
The RIXS intensities above the horizontal lines (white) are magnified by a factor indicated in panels.}
\label{fig_app1}
\end{figure}

Figure~\ref{fig_app1} shows the calculated $L_3$-RIXS map of NiO, in which a charge-transfer channel from the x-ray excited Ni ion and the conduction states above $E_F$ is eliminated in the RIXS process.
In practice, the $V(\varepsilon)$ intensities are set to zero for $\varepsilon > 0$~eV by hand in the whole RIXS calculation.
The $V(\varepsilon)$ above $E_F$ mainly corresponds to the hybridization with the UH states outside the impurity Ni site.
Thus the unbound EHP excitations with the UH states outside the excited Ni site are forbidden in the resultant spectra. 

\bibliography{rixs_dmft}

\end{document}